\documentclass[aps,pre,showpacs,showkeys,preprint,groupedaddress]{revtex4}
\usepackage{graphicx}
\usepackage{dcolumn}
\usepackage{bm}

\begin{document}

\title{Angular momentum induced phase transition in spherical gravitational
systems: N-body simulations}
\author{Peter Klinko, Bruce N. Miller}
\affiliation{Department of Physics, Texas Christian University, \\
Fort Worth, Texas 76129}
\date{\today}

\begin{abstract}
The role of thermodynamics in the evolution of systems evolving under purely
gravitational forces is not completely established. Both the infinite range
and singularity in the Newtonian force law preclude the use of standard
techniques. However, astronomical observations of globular clusters suggests
that they may exist in distinct thermodynamic phases. Here, using dynamical
simulation, we investigate a model gravitational system which exhibits a
phase transition in the mean field limit. The system consists of rotating,
concentric, mass shells of fixed angular momentum magnitude and shares
identical equilibrium properties with a three dimensional point mass system
satisfying the same condition. The mean field results show that a global
entropy maximum exists for the model, and a first order phase transition
takes place between "quasi-uniform" and ''core-halo'' states, in both the
microcanonical and canonical ensembles. Here we investigate the evolution
and, with time averaging, the equilibrium properties of the isolated system.
Simulations were carried out in the transition region, at the critical
point, and in each clearly defined thermodynamic phase, and striking
differences were found in each case. We find full agreement with mean field
theory when finite size scaling is accounted for. In addition, we find that
(1) equilibration obeys power law behavior, (2) virialization,
equilibration, and the decay of correlations in both position and time, are
very slow in the transition region, suggesting the system is also spending
time in the metastable phase, (3) there is strong evidence of long-lived,
collective, oscillations in the supercritical region.
\end{abstract}


\maketitle

\section{Introduction}

In contrast with ''normal'' systems with short range interactions, the
thermodynamics of self-gravitating systems is non-extensive and, because of
the infinite range and singularity of the Newtonian potential, cannot be
treated by standard methods. A partial remedy for these problems can be
constructed by confining the system in a finite volume and either using a
regularized Newtonian pair interaction potential, or considering the N-body
system in the mean-field limit where it is possible to construct an analytic
theory of $f(\mathbf{r},\mathbf{v})$ , the single particle density in
position and velocity. The first mean-field formulations showed that (1)
spherically symmetric density profiles represent the states of highest
entropy and (2) a global entropy maximum does not exist \cite{antonov}. It
is always possible to increase the entropy by simultaneously increasing the
central density and transferring mass to a diffuse ''halo'' to control the
value of the energy. This phenomenon, called gravothermal catastrophe in the
literature, reflects the fact that an isolated and bounded gravitational
system cannot be in equilibrium in the mean-field limit. Locally stable and
unstable entropy extrema can exist, however, if the system's energy is above
a critical value \cite{antonov,lyndbw}, and their stability has been
investigated by several authors \cite{padna1,katz1,katz2}.

In addition to the total energy, $E$, in the mean field limit the sum of
squares of the angular momentum, $L_{2}=\lim\limits_{N->\infty }1/N\sum
l_{i}^{2}$ , where $\mathbf{\ l}_{i}$ is the angular momentum of a system
element, is also an integral of motion for a spherically symmetric
gravitational system, \cite{binney} . If, along with the energy, this
constraint is also included in MFT (mean field theory) the result is an
anisotropic distribution in velocity. The result was first obtained by
Eddington from a different route \cite{Eddington}. Until recently, this
second integral has been ignored in investigations of thermodynamic
stability in the mean field (Vlasov) limit, leaving open the possibility
that a centrifugal barrier could prevent collapse and stabilize the system.
In fact, we have recently shown that even its inclusion in MFT cannot
resolve the gravothermal catastrophe, and it persists in both the
generalized microcanonical ($E-L_{2} $) and canonical ($\beta -\gamma $)
ensembles \cite{klinkmillL2}. The extremal solutions have the form $f\propto
\exp (-\beta \epsilon )\exp (-\gamma l^{2})$, which coincides with the well
known anisotropic density fit models that have been applied to globular
cluster observations with some success(e.g. King-Michie models) \cite
{gunn,Heggie}.

In their seminal work on the gravothermal catastrophe, Lynden-Bell and Wood 
\cite{lyndbw} pointed out that in the absence of the short range
singularity, the possibility existed for a gravitational system to exist in
different thermodynamics phases. Using mean-field models with a regularized
Newtonian potential which has been dissected to remove the singularity \cite
{kiessling1} or, equivalently, introducing repulsion at short-range by
imposing a local equation of state in the mean field picture \cite
{aronson,stahlkiessling2,padna1}, it has been shown that a first order phase
transition can occur in both the microcanonical (MCE) and canonical (CE)
ensembles. Unfortunately, for finite $N$ systems, there are no exact
microcanonical results available which allow the rigorous proof of a phase
transition or catastrophe. However, in the canonical case, it has been
rigorously proved that the system of gravitating point masses is in a
collapsed state in equilibrium in the absence of regularization \cite
{kiessling1}. Moreover, Monte-Carlo simulations for a regularized Newtonian
potential confirm the gravitational phase transition in CE \cite{sanchez} in
large $N$ systems.

While mean field theories support the existence of phase transitions in
gravitational systems, it is important to point out that there is no
guarantee that these equilibrium states will be realized by dynamical
evolution. In fact, there is no proof that the two operations of taking (1)
the mean field limit, or (2) the infinite time average, commute \cite{latora}%
. Rather, simulations of the one dimensional self-gravitating system
consisting of parallel mass sheets provide strong evidence to the contrary 
\cite{mineufeix}. As a consequence, the relation between maximum entropy
solutions of the stationary mean field equation and the time average
distribution functions resulting from N body simulation, or dynamical
evolution in nature, has not been fully established. This is a deep question
which will not be explored further here. Thus, although much is known
concerning the ''equilibrium'' properties in the mean field limit, the
dynamical properties of gravitational phase transitions are not well known
due to a lack of true N-body simulations, which are also important for
explaining the evolution of stellar clusters, galaxies, etc.. At the present
time the mean-field predictions of the gravitational phase transition have
only been dynamically confirmed (in both MCE and CE) for the model system
consisting of irrotational, concentric, mass shells \cite{millyounglett}. In
that model, the Newtonian singularity was screened by the introduction of an
inner barrier which excluded mass from the system center.

The aim of this paper is to investigate and understand the dynamical
features of gravitational phase transitions in N-body simulations for the
model of a purely Newtonian system in which $l_{i}^{2}=l^{2}=L_{2}$ for each
system element. We will refer to this system as the $l^{2}$ model. We will
explicitly investigate a system of rotating, concentric, mass shells. In the
mean field limit, this system shares important features, e.g. the
equilibrium density and radial velocity distribution, with the more
realistic system of point masses. Recently, with I. Prokhorenkov, we showed
that the gravitational phase transition is present in both MCE and CE in
certain regions of $l^{2}$ and we studied its properties using mean field
theory \cite{klinkmillprok}. Moreover, we rigorously proved the existence of
an upper bound for the entropy in the MCE (lower bound in the free energy in
the CE) for $l^{2}\neq 0$ in the same work. It is important to understand
that this model demonstrates the significance of the influence of angular
momentum on the thermodynamics of self-gravitating systems, even if they are
spherical. The generalized microcanonical ($E-L_{2}$) and canonical ($\beta
-\gamma $) ensembles discussed earlier \cite{klinkmillL2}are appropriate for
large $N$ spherical systems where angular momentum exchange occurs. While
providing the most general mean field description \cite{klinkmillL2}, $L_{2}$
is still not a sufficient constraint to resolve the gravothermal
catastrophe. Clearly the culprit is angular momentum exchange, which is
still permitted even if both energy and $L_{2}$ are fixed, and allows the
transfer of mass to the system center. An open question is the possible
existence of additional mechanisms for establishing a centrifugal barrier
which prevents, or strongly inhibits, collapse in nature, e.g. in globular
clusters or molecular clouds. This will be taken up in the final section.

In the following, first we briefly review the $l^{2}$ model, including the
mean-field predictions, and discuss the main features of the N-body
algorithm we designed for its dynamical simulation. Next we turn to the
simulation results in a region of the $(E,l^{2})$ phase plane containing the
microcanonical phase transition region. In each phase we compare the
time-averaged equilibrium properties with the predictions of mean field
theory resulting from our earlier investigation, and also touch on finite
size effects. We then go on to study both equilibrium and dynamical features
which cannot be predicted by mean field theory, such as the variance of
fluctuations, and correlations in both time and position. In addition to the
system behavior near the phase transition, we pay particular attention to
the critical point, and the supercritical region. Finally we consider the
surprising features exhibited by different stages of the relaxation process
itself, and their dependence on energy, $l^{2}$, and population, and discuss
the possible presence of collective modes.

\begin{figure}[tbp]
\centerline{\includegraphics[scale=0.5]{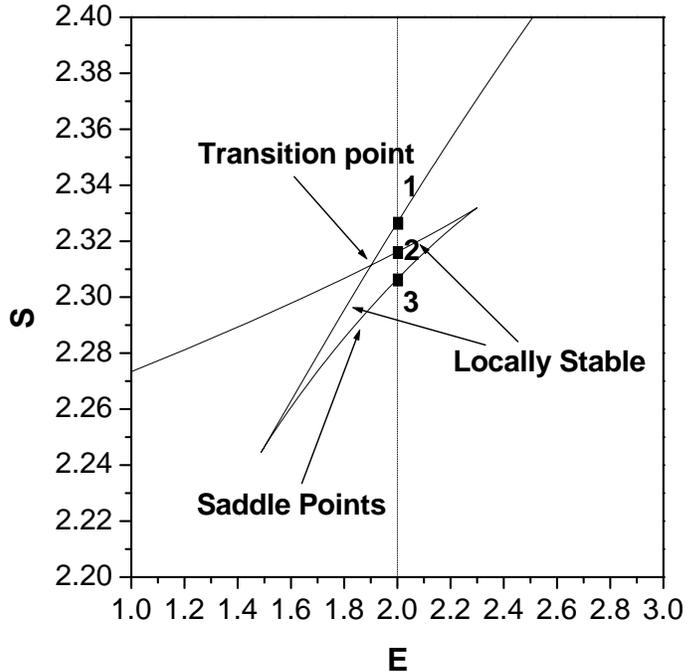}}
\caption{Plot of the entropy extrema in the microcanonical ensemble for 
$l^{2}=5\times 10^{-5}$ in the mean-field limit. As we can see, in the
transition region multiple solutions are present, and we marked the
solutions for $E=2$ with labels 1, 2, and 3. Only phase 1 (quasi-uniform
phase) is a global entropy maximum above the transition point ($E=1.9)$,
while phase 2 (condensed phase) is locally stable, and phase 3 is a saddle
point. }
\label{figtransmean}
\end{figure}

\begin{figure}[tbp]
\centerline{\includegraphics[scale=0.5]{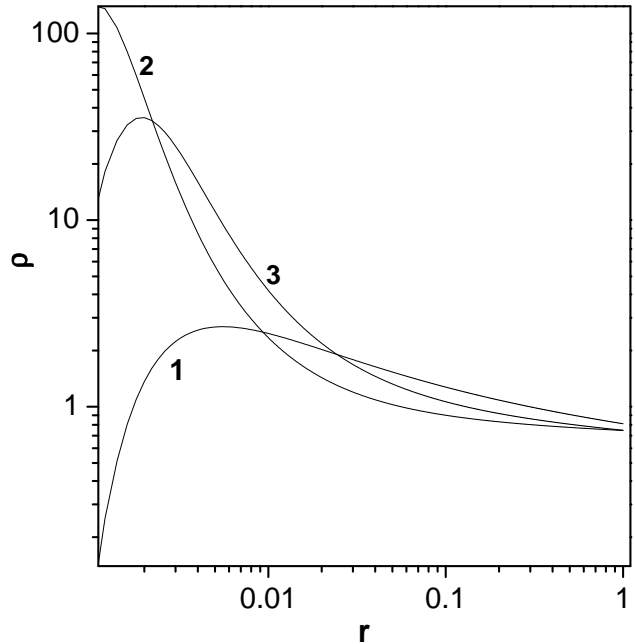}}
\caption{Density profiles of the three distinct phases labelled in Fig.~\ref
{figtransmean} for $E=2$ in the transition region. Above the transition
point only the quasi-uniform phase (phase 1) is globally stable, while the
condensed phase (phase 2) is locally stable. The third entropy extremum
(phase 3) is unstable.}
\label{figdensityprof}
\end{figure}

\begin{figure}[tbp]
\centerline{\includegraphics[scale=0.5]{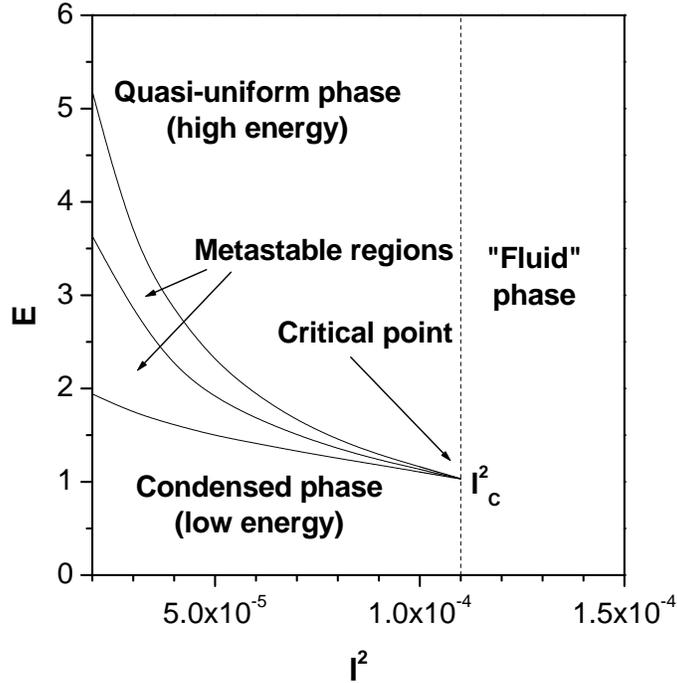}}
\caption{The mean-field microcanonical phase diagram. The system can exist
in 3 different types of phase depending on the energy and $l^{2}$, and we
named these phases in analogy with normal systems. We also indicated the
metastable regions which are important for understanding the dynamical
behavior of the system near the transition point.}
\label{figphasediagram}
\end{figure}

\begin{figure}[tbp]
\centerline{\includegraphics[scale=0.5]{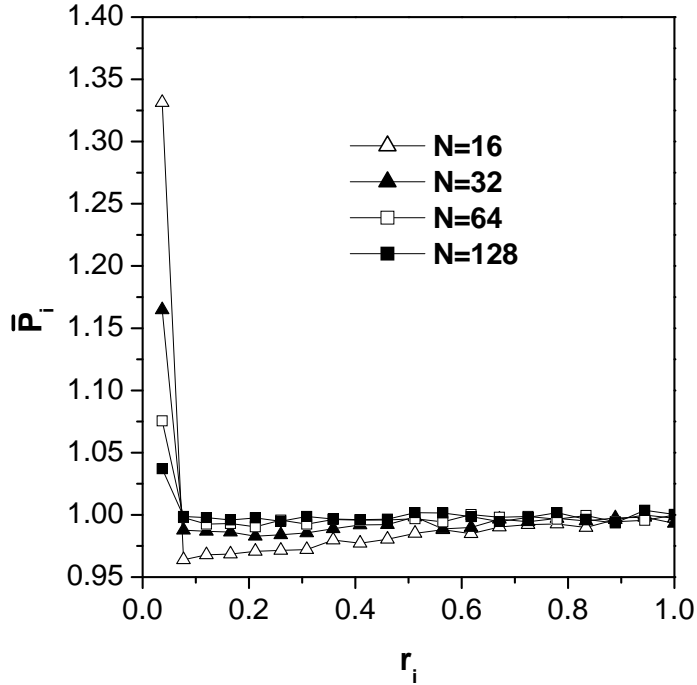}}
\caption{The time averaged relative populations $\bar{P}_{i}$ of relaxed
systems in the high energy region at $E=4$ and $l^{2}=5\times 10^{-5}$ with $%
N_{bin}=20$. The bins were obtained from the equal-mass radii of the
mean-field equilibrium density profile. We observe good convergence to the
mean-field density profile with increasing $N$. As a result of finite size
effects, the central density is higher than the mean-field density profile
in the high energy region, and causes a reduction in the virial ratio.}
\label{figdensityhigh}
\end{figure}

\begin{figure}[tbp]
\centerline{\includegraphics[scale=0.5]{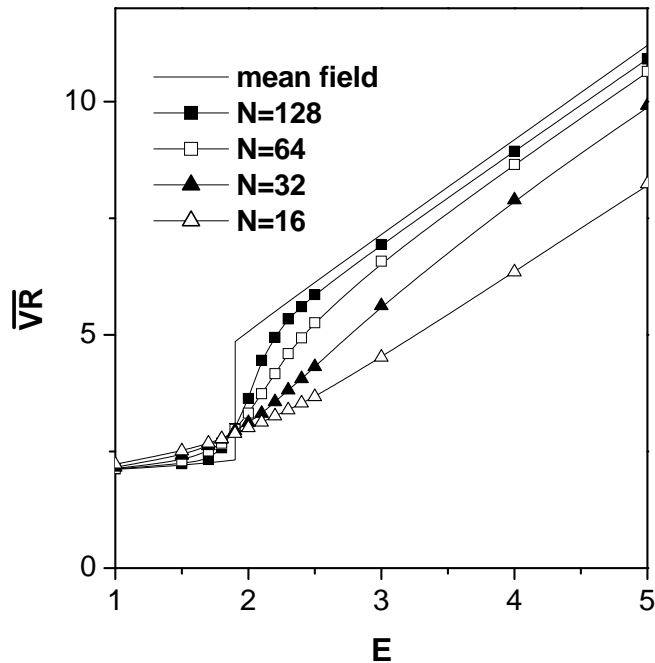}}
\caption{The averaged virial ratio $\overline{VR}$ for $56$ simulations for $%
N=16, 32, 64, 128$ in and around the mean-field transition region. We also
included the mean-field results for the globally stable states which show
that the system undergoes a first order phase transition at $E=1.9$, where
the virial ratio becomes discontinuous. The simulation results converge to
the mean-field predictions with increasing $N$, but the transition point is
shifted and the transition region is rounded, in good agreement with
finite-size scaling.}
\label{figphasesim}
\end{figure}

\begin{figure}[tbp]
\centerline{\includegraphics[scale=0.5]{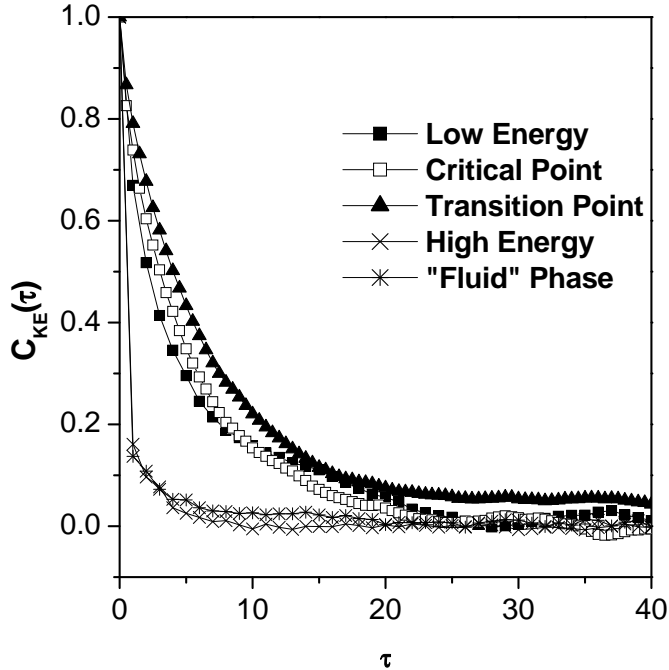}}
\caption{The time correlation function of the kinetic energy $C_{KE}\left( 
\protect\tau \right) $ in 5 different phase regions for $N=64$, where $%
\protect\tau $ is in units of $t_{dyn}$. In both the high energy ($%
E=4,l^{2}=5\times 10^{-5} $) and supercritical phase ($E=4,l^{2}=5\times
10^{-3}$), correlations are smaller, while in the low energy phase ($%
E=1.5,l^{2}=5\times 10^{-5}$), at the transition point ($E=1.9,l^{2}=5\times
10^{-5}$), and at the critical point ($E=1.052,l^{2}=1.1\times 10^{-4}$)
relatively stronger correlations can be observed. Note that only the initial
decay of the correlation functions follows an exponential behavior, then a
long tail develops.}
\label{figKEcorr}
\end{figure}

\begin{figure}[tbp]
\centerline{\includegraphics[scale=0.5]{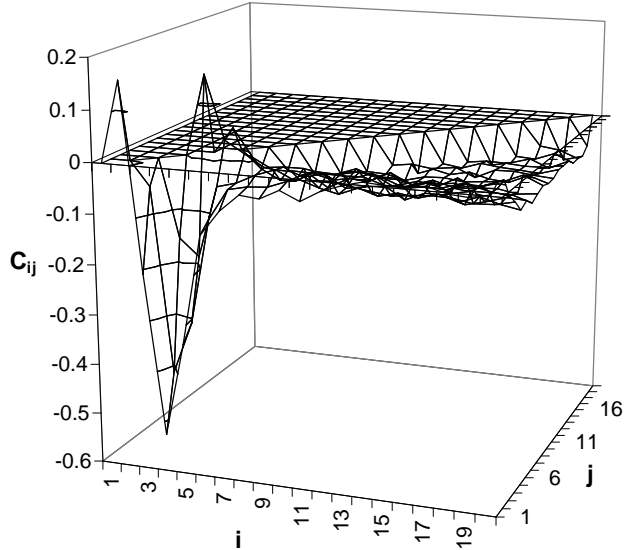}}
\caption{Position correlation matrix in relative bin populations $C_{ij}$,
for $N_{bin}=20$ and $i>j$ in the low energy phase ($E=1,l^{2}=5\times
10^{-5},N=64$). The bin radii represent the equal-mass layers obtained from
the mean-field density profile. Strong anti-correlation is present close to
the high density central core region.}
\label{figpopcorrlow}
\end{figure}

\begin{figure}[tbp]
\centerline{\includegraphics[scale=0.5]{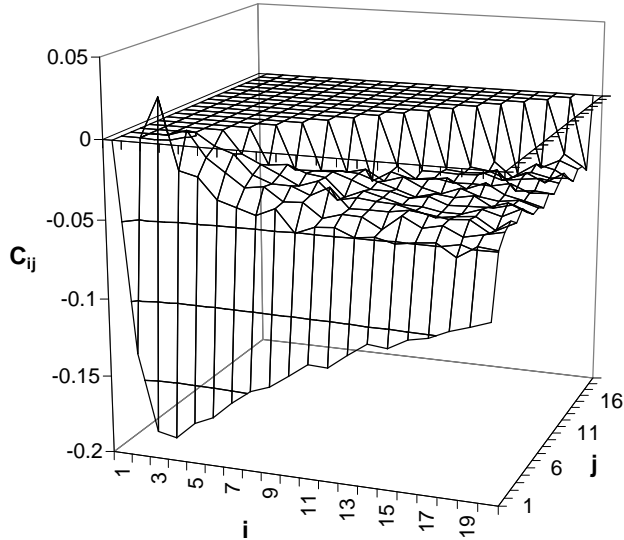}}
\caption{Position correlation matrix in relative bin populations $C_{ij}$,
for $N_{bin}=20$ and $i>j$ in the phase transition region ($%
E=2,l^{2}=5\times 10^{-5},N=64$). The bin radii represent the equal-mass
shells obtained from the mean-field density profile of phase 1 in Fig.~\ref
{figtransmean}. Strong anti-correlation is present between the center and
the remainder of the system.}
\label{figpopcorrphase}
\end{figure}

\begin{figure}[tbp]
\centerline{\includegraphics[scale=0.5]{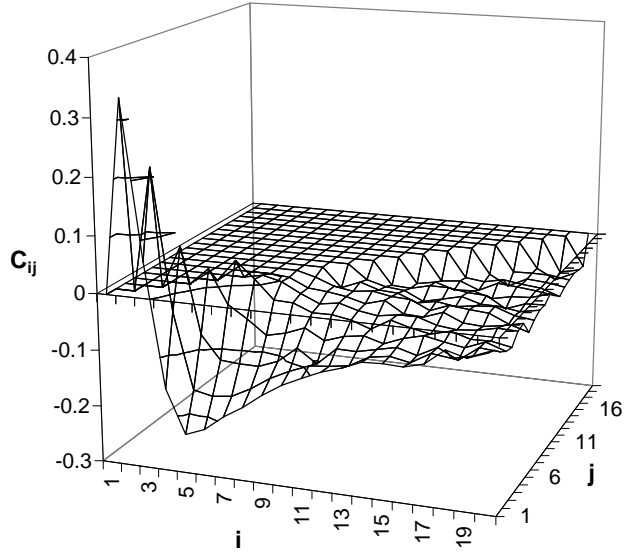}}
\caption{Position correlation matrix in relative populations $C_{ij}$, for $%
N_{bin}=20$ and $i>j$ near the critical point of the phase diagram ($%
E=1.052,l^{2}=1.1\times 10^{-4},N=64$). The bin radii represent the
equal-mass shells obtained from the mean-field density profile. The behavior
is similar to that of the low energy phase, although the anti-correlating
region is wider. This reflects the fact that, at the critical point, the
difference between the two distinct phases vanishes. This intermediate state
still has a core-halo structure, but with a more extended core region.}
\label{figpopcorrcrit}
\end{figure}

\begin{figure}[tbp]
\centerline{\includegraphics[scale=0.5]{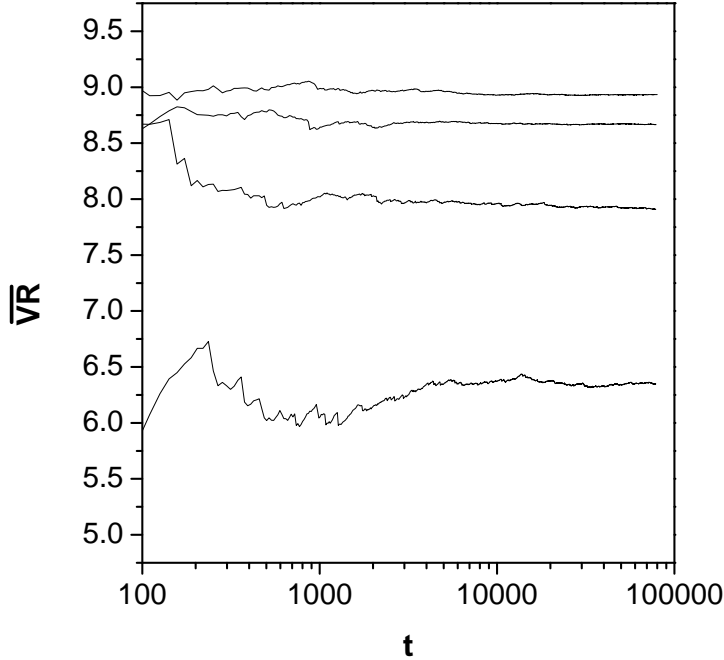}}
\caption{Violent relaxation in the high energy phase ($E=4,l^{2}=5\times
10^{-5}$) in four simulations with $N=16, 32, 64 , 128$. In the figure $N$
is increasing from bottom to top. We observe that virialization is faster
with increasing $N$. For $N=64, 128$ the differences become marginal as we
approach the mean-field limit.}
\label{figvirrelaxhigh}
\end{figure}

\begin{figure}[tbp]
\centerline{\includegraphics[scale=0.5]{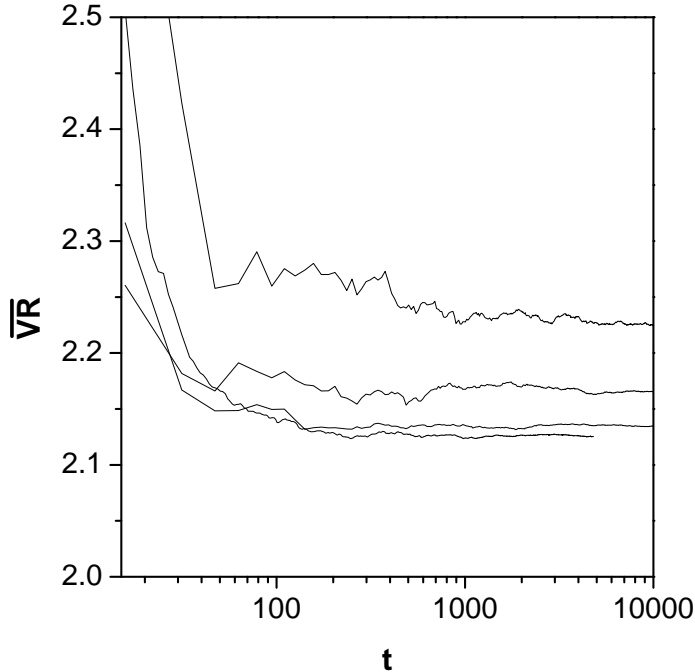}}
\caption{Violent relaxation in the low energy phase ($E=1,l^{2}=5\times
10^{-5}$) at $N=16, 32, 64, 128$ (from top to bottom in the figure). Here
the virialization is also fast for larger values of $N$, and takes place on
a much shorter time-scale than in the high energy phase. Note that for $%
N=64, 128$, the system virializes in $100t_{dyn}$. }
\label{figvirrelaxlow}
\end{figure}

\begin{figure}[tbp]
\centerline{\includegraphics[scale=0.5]{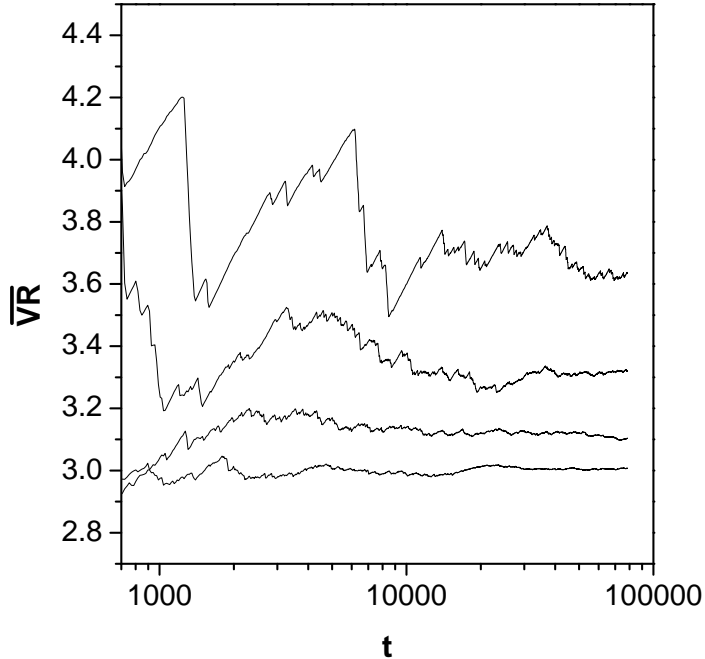}}
\caption{Violent relaxation in the phase transition region ($%
E=2,l^{2}=5\times 10^{-5} $) in four different simulations. From bottom to
top $N=16, 32, 64, 128$. Compared with the high and low energy phases, the
relaxation is slower for larger $N$ as opposed to the process in the high
and low energy phases. As we approach the mean-field limit, the phase
transition becomes sharper and the increasing influence of the meatstable
phase slows down the relaxation.}
\label{figvirrelaxphase}
\end{figure}

\begin{figure}[tbp]
\centerline{\includegraphics[scale=0.5]{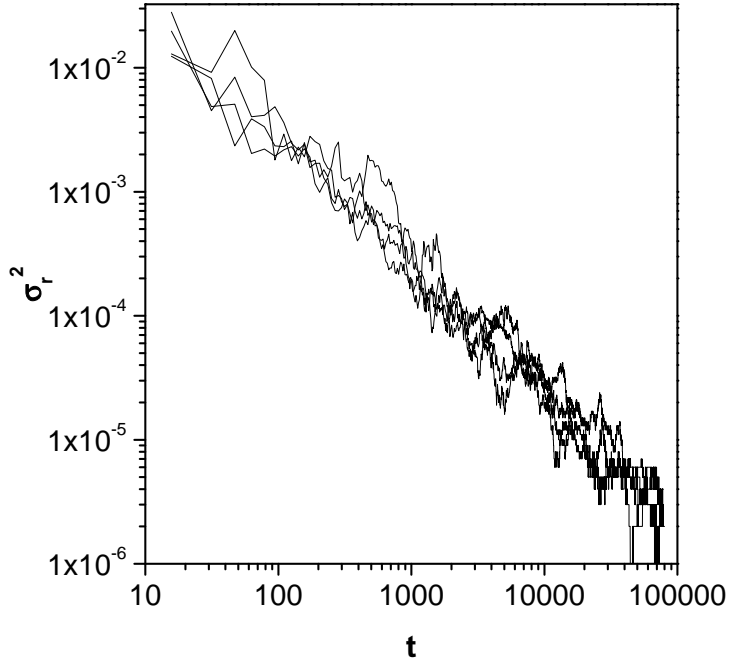}}
\caption{Plots of $\protect\sigma _{r}^{2}$ vs. time which indicate the
relaxation process in $\protect\mu $ space for simulations with $N=16, 32,
64, 128$ in the high energy phase ($E=4,l^{2}=5\times 10^{-5}$). In order to
avoid statistical error while sampling the populations, we chose snapshot
times $t_{s}=0.125, 0.25, 0.5, 1$ for the corresponding values of $N$. For
all $N$, the same power-law type of relaxation, $\protect\sigma %
_{r}^{2}\propto 1/t$, can be observed.}
\label{figrelaxhigh}
\end{figure}

\begin{figure}[tbp]
\centerline{\includegraphics[scale=0.5]{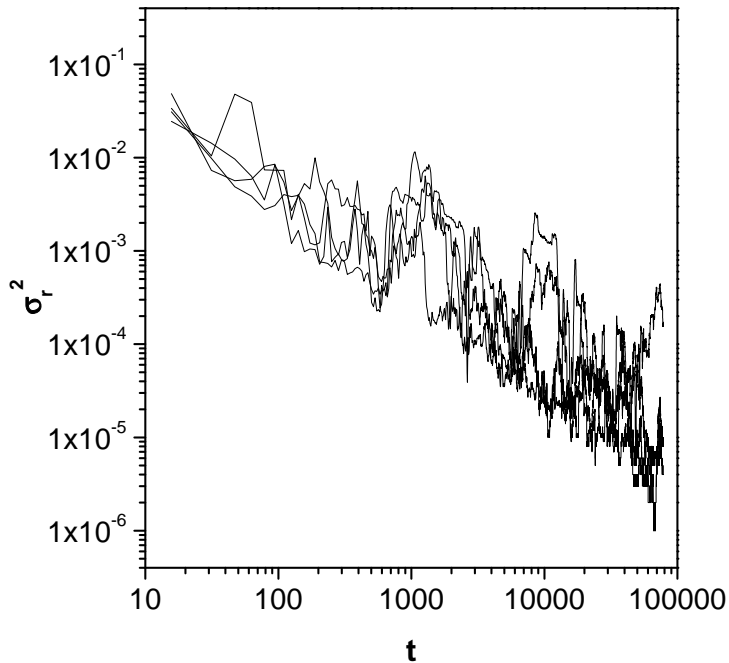}}
\caption{Plots of $\protect\sigma _{r}^{2}$ vs. time which indicate the
relaxation process in $\protect\mu $ space with simulations $N=16, 32, 64,
128$ in the phase transition region ($E=2,l^{2}=5\times 10^{-5}$). In order
to avoid statistical error during sampling the populations, we chose
snapshot times $t_{s}=0.125, 0.25, 0.5, 1$ for the corresponding $N$. The
power-law behavior can still be observed, but large fluctuations start to
take place with increasing $N$. As a result of the influence of the
metastable phase, relaxation is slower for larger $N$ in this region.}
\label{figrelaxphase}
\end{figure}

\begin{figure}[tbp]
\centerline{\includegraphics[scale=0.5]{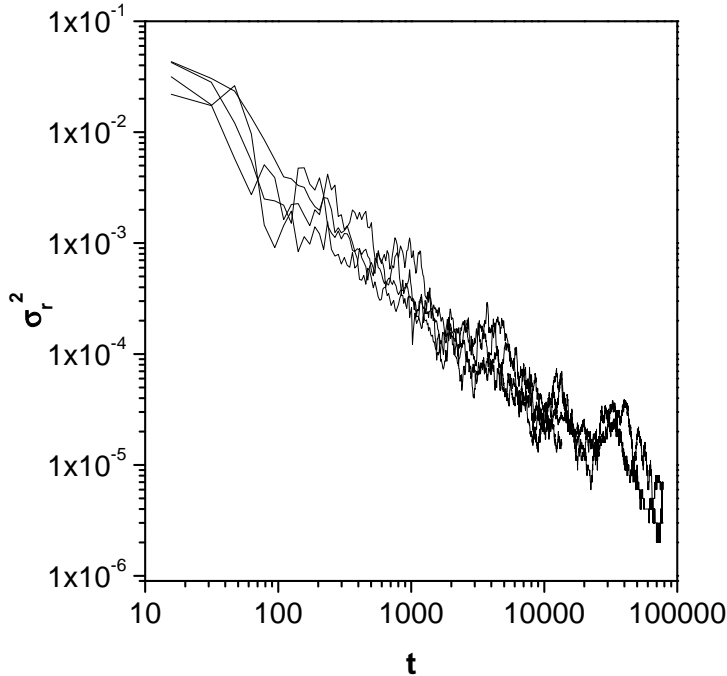}}
\caption{Plots of $\protect\sigma _{r}^{2}$ vs. time which indicate the
process of relaxation in $\protect\mu $ space for simulations with $N=16,
32, 64, 128$ in the low energy phase ($E=1,l^{2}=5\times 10^{-5}$). In order
to avoid statistical error during sampling the populations, we chose
snapshot times $t_{s}=0.125, 0.25, 0.5, 1$ for the corresponding values of $%
N $'s. In the high energy phase, we observe a similar power-law relaxation
which also follows $\protect\sigma _{r}^{2}\propto 1/t$.}
\label{figrelaxlow}
\end{figure}

\begin{figure}[tbp]
\centerline{\includegraphics[scale=0.5]{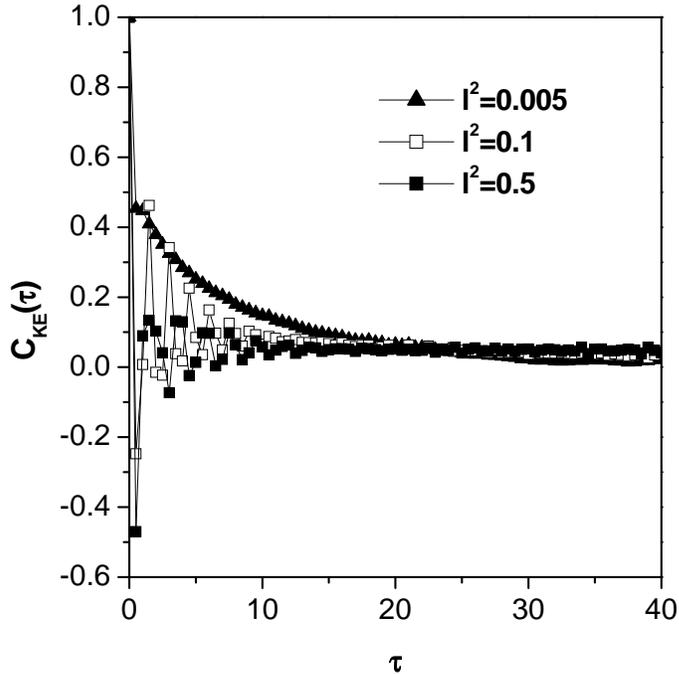}}
\caption{The time correlation of the kinetic energy $C_{KE}\left( \protect%
\tau \right) $ versus time for three regions of the supercritical phase at
the energy $E=0$. With increasing $l^{2}$, an oscillating part starts to sit
on top of the correlation functions which suppresses the early exponential
behavior, and a very long tail starts to develop.}
\label{figKEcorrfluidl2}
\end{figure}

\begin{figure}[tbp]
\centerline{\includegraphics[scale=0.5]{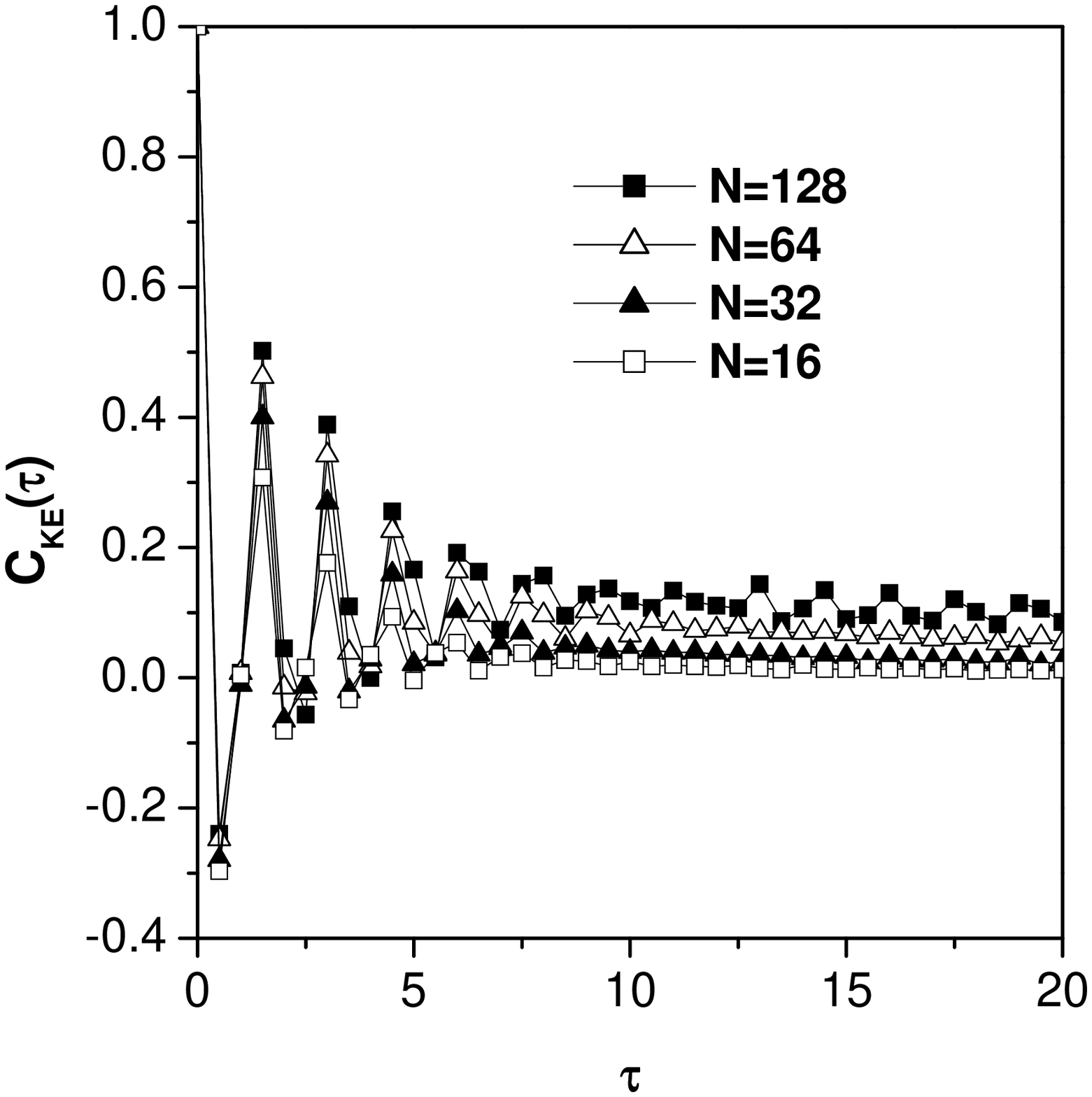}}
\caption{$N$ dependence of the oscillations in $C_{KE}\left( \protect\tau %
\right) $ at $E=0$ and $l^{2}=0.1$. As we approach the mean-field limit,
oscillations persist for a longer time and correlations in the tail grow.}
\label{figKEcorrfluidN}
\end{figure}

\section{The $l^{2}$ model and mean-field results}

The mean field, or Vlasov, limit is obtained by letting $N\rightarrow \infty 
$ while controlling both the total mass and energy \cite{binney, braunhepp,
klinkmillL2}. Taking the limit results in a nonlinear partial differential
equation, the Vlasov equation, for the evolution of $f(\mathbf{r},\mathbf{v}%
) $ which is first order in the time. In contrast with the Boltzmann
equation, there is no collision term and, consequently, the system lacks an
increasing entropy as time progresses. Nonetheless, it is possible to
construct maximum entropy solutions of the stationary system for qualifying
systems. For the special case of spherical symmetry, the problem reduces to
a pair of coupled, nonlinear, differential equations for the local density
which can be integrated numerically \cite
{chandra,binney,klinkmillprok,klinkmillL2}.

Here we introduce the $l^{2}$ model of a spherically symmetric distribution
of self-gravitating particles in the mean-field limit confined in a finite
radius $r\leq b$. A single ''test'' particle in the system has the
Hamiltonian per unit mass

\begin{equation}
H=\frac{1}{2}v^{2}+\frac{l^{2}}{2r^{2}}+\Phi \left( r\right) ,
\label{l2hamdens}
\end{equation}
where $l^{2}$ is fixed, $\Phi (r)$ is the gravitational potential, and we
chose units where $M=b=1$ . The thermodynamics of this model has been worked
out rigorously in both MCE and CE \cite{klinkmillprok}. Since the dynamical
system is effectively one dimensional, $f(r,v)$ now only depends on the
radial coordinate $r$ and radial velocity $v.$ The Shannon entropy, $S[f],$
is a functional of $f$, and is simply expressed by 
\[
S[f]=-\int_{-\infty }^{\infty }\int_{0}^{b}f\ln fdrdv 
\]
where we have chosen unity for the natural measure in the $(r,v)$ plane, and
units where the Boltzmann constant, $k_{B}=1$. As a result of the angular
momentum constraint, the global maximum of the entropy (or minimum of the
free energy in CE) exists for all $l^{2}\neq 0$, and first order phase
transitions are predicted both in MCE and CE for a specific range of values
of $l^{2}$ \cite{klinkmillprok}. As expected, extrema of $S[f]$ with respect
to the constraints of normalization and the total energy 
\[
E=\int_{-\infty }^{\infty }\int_{0}^{b}f\left( \frac{1}{2}v^{2}+\frac{l^{2}}{%
2r^{2}}+\frac{\Phi }{2}\right) drdv, 
\]
occur when $f\sim \exp (-\beta H)$.

For an isolated system, global thermodynamic stability is determined by the
state of maximum entropy. In our earlier work \cite{klinkmillprok} we showed
that for $l^{2}>l_{c}^{2}\cong 1.1\times 10^{-4}$, the system can only exist
in a single phase, while for $l^{2}<l_{c}^{2}$ two stable phases are
available, depending on the energy. In Figure \ref{figtransmean}, we present
the extremal entropy solutions (including both globally and locally stable
maxima, and the saddle points in the transition region) for $l^{2}=5\times
10^{-5}$ in the MCE. In Fig.~\ref{figdensityprof}, we plot the linear
density profiles for each of the three phases for $E=2$ (Fig.~\ref
{figtransmean}). In the transition region and above the transition point at $%
E=2$, only phase 1 is globally stable (quasi-uniform phase), phase 2 is only
locally stable (condensed phase), while phase 3 is a saddle point. We can
clearly see from the figures that at about $E=1.9$, a microcanonical phase
transition takes place between a quasi-uniform and a centrally dense
core-halo state. This type of phase transition is a unique feature of a
self-gravitating system, since the two different phases cannot coexist. The
selected value of $l^{2}$ is sufficiently small that the stable and
metastable phases are well separated, yet not so small that dynamical
simulation becomes intractable.

It is important to point out that, for gravitational systems, the CE and MCE
formulations are not equivalent. For example, the value of $l^{2}$ at the
critical point is different in each ensemble, and the transition region in
MCE is unstable in CE. This has been discussed in detail elsewhere, and we
will not pursue details here \cite{stahlkiessling2,padna2}. In Figure \ref
{figphasediagram}, we present the micocanonical phase diagram for the model
where, in addition to the coexistence curve, we also indicate the boundaries
of regions where a second, metastable, phase exists. We easily observe that
there is a critical value of $l^{2},$ where the width of the metastable
region vanishes and beyond which a transition doesn't occur. This is a true
critical point: Keeping in mind the analogy with the liquid-vapor
transition, states with $l^{2}>l_{c}^{2}$ which characterize the
supercritical region are analogous to the fluid phase. A useful
representative of both the low and high energy phase is the virial ratio $%
VR=2KE/\left| PE\right| $, where $KE$ and $PE$ are, respectively, the total
system kinetic and potential energy. $VR$ is discontinuous at the transition
point (also see Figure \ref{figphasesim}) and will play the roll of order
parameter in what follows.

\section{N-body code and initial conditions}

\label{Nbody}

Our N-body code models a system of concentric, rotating, infinitesimally
thin, spherical mass shells confined in a finite radius $b$, where each
shell can rotate about any axis. In this system, the $i^{th}$ shell has the
following Hamiltonian per unit mass, 
\begin{equation}
\frac{H_{i}}{m_{i}}=\frac{1}{2}v_{i}^{2}-\frac{m_{i}/2+\sum_{k=1}^{i-1}m_{k}%
}{r_{i}}+\frac{l_{i}^{2}}{2Ir_{i}^{2}}.  \label{shellham}
\end{equation}
For a thin shell $I=2/3$, and the motion of each shell is integrable between
crossings. The main advantages of the model over the conventional $N$ body
system of point masses are that it preserves spherical symmetry, and it is
only necessary to evaluate those discrete time events which occur either
when two shells cross one another, or one shell arrives at a turning point
or the outer boundary. Between these events, the equations of motion induced
by $H_{i\text{ }}$ are easily integrated in closed form yielding the time as
an explicit function of position. Thus the amount of numerical error and
computation time can be greatly reduced, and simulations can be performed
until thermal equilibrium is obtained. Another advantage of the model is
that, for sufficiently large N, it approaches an exact description of a
spherical Newtonian N-body point mass system. This can be seen in Eq.~(\ref
{l2hamdens}) by recognizing that, with the exception of the value of $I$,
both the equations of motion for the radial coordinate, and the coupled
first order equations for the equilibrium density, are the same. Clearly $%
I=1.0$ for the point mass system. Thus if we let $\mathbf{l}_{i}^{P}=\mathbf{%
\ l}_{i}^{S}/I$ $,$ \ where $\mathbf{l}_{i}^{P}$ and $\mathbf{l}_{i}^{S}$
are, respectively, the angular momentum per unit mass of the $i^{th}$ point
mass and the corresponding shell, we can establish a correspondence between
the two systems. Keeping this connection in mind, we can compare the
dynamical simulations of rotating concentric shells presented here with our
recent theoretical, mean field, study of a point mass system for the $l^{2}$
model.

To compare our results with a system of point particles with fixed $l^{2}$,
here we prepared our shell system with the equivalent angular momenta of
fixed magnitude $2l^{2}/3$, initial positions uniformly distributed in the
interval $[0,b]$, and initial radial velocities randomly oriented with fixed
magnitude. In order to maintain consistency with the mean-field results \cite
{klinkmillprok}, reduce numerical errors, and preserve the system of units
applied in the mean field model, we used similar units with $G=b=1$ and $%
m_{i}=1/N.$ We took snapshots of the complete system state after the passing
of each dynamical time $t_{dyn}=\pi /2$ \cite{binney} the simulation when
the system was well relaxed. For the definition of relaxation, we assumed
that the system has equilibrated if it has explored the available phase
space and, on average, the one particle probability density function has
converged. To quantify convergence, we divided the radius into a fixed
number of bins, $N_{bin}$, in which each shell has equal probability of
occurrence in accordance with the mean-field equilibrium probability density
profile for the given energy and $l^{2}$. We define the system to be relaxed
in $t=2kt_{dyn}$ if, for a suitable $\delta >0$ (typically $10^{-7}$ in our
simulations), 
\begin{equation}
\sigma _{r}^{2}\left( 2k\right) =\frac{1}{N_{BIN}}\sum_{i=1}^{N_{BIN}}\left( 
\bar{P}_{i}\left( k\right) -\bar{P}_{i}\left( 2k\right) \right) ^{2}<\delta
\label{relax}
\end{equation}
where $\bar{P}_{i}\left( n\right) =\sum_{k=1}^{n}N_{i}\left( k\right)
N_{BIN}/(Nn)$ is the time averaged relative population in cell $i$ and $%
N_{i}\left( k\right) $ is the number of shells in bin $i$ at the time $%
kt_{dyn}$ . Other important statistical properties of the system include
fluctuations at fixed time, and correlations in time and position.

To investigate the approach to mean field (or Vlasov) behavior, we directly
computed the time averaged variance in kinetic energy from simulations, as
well as the variance of the population of each bin, for selected points in
the $(E,$ $l^{2})$ phase plane. In each case, their dependence on system
population was carefully studied. To obtain selected information about the
decay of fluctuations in time, we studied the correlation of the kinetic
energy in time, 
\begin{equation}
C_{KE}\left( \tau \right) =\frac{1}{\left( n-\tau \right) \sigma _{KE}^{2}}%
\sum_{k=1}^{n-\tau }\left( KE\left( k+\tau \right) -\bar{KE}\left( n\right)
\right) \left( KE\left( k\right) -\bar{KE}\left( n\right) \right) .
\label{KEcorr}
\end{equation}
where $\sigma _{KE}^{2}$ is the variance of the system's kinetic energy. To
gain information about the range of correlation in position, we computed the
correlation matrix between each pair of bins for the relative populations $%
P_{i}\left( t\right) =N_{i}\left( k\right) N_{BIN}/N:$ 
\begin{equation}
C_{ij}=\frac{1}{n\sigma _{i}^{P}\sigma _{j}^{P}}\sum_{k=1}^{n}\left(
P_{i}\left( k\right) -\bar{P}_{i}\left( n\right) \right) \left( P_{j}\left(
k\right) -\bar{P}_{j}\left( n\right) \right) .  \label{popcorr}
\end{equation}
where $\sigma _{i}^{P}$ is the standard deviation of the relative population
in cell $i$.

\section{Microcanonical phase transition}

In order to investigate the agreement of dynamical simulations with the
predictions of MFT discussed above, we carried out simulations for four
system populations, $N=16,32,64,128$. While particular attention was focused
on the phase transition region of $l^{2}=5\times 10^{-5}$, the system
dynamics was also investigated in each of the thermodynamic phases in the $%
(E,l^{2})$ plane predicted by mean field theory. Although it is possible to
study evolution in systems with larger particle numbers, substantially more
CPU time is required, and it becomes increasingly more difficult to
numerically resolve successive shell crossings which may become closely
spaced in time and position during the course of long simulations. As in the
case of the irrotational shells studied earlier, \cite{milleryoungchaos} the
dynamics of the system showed highly chaotic behavior and substantial mixing
in $\mu $ space, i.e. in the $(r,v)$ plane. However, as we shall see later,
the rate of mixing strongly depends on the thermodynamic phase in which the
system is prepared.

We selected the time averaged bin populations as useful measures of
agreement between the predictions of MFT and actual dynamical simulations.
Outside the transition region, the relaxed density profiles converged to the
corresponding mean-field density prediction with increasing $N.$ This is
apparent in Fig.~\ref{figdensityhigh} which shows the averaged relative
populations for $N_{bin}=20$ and $E=4$. It is clear from the figure that
convergence in the innermost cell was much weaker than in the remainder, but
improves with increasing $N$. We noticed that, as a rule, both convergence
to the Vlasov limit with increasing $N$ and the evolution of the system in
time to the equilibrium state were slower in the phase transition region
than in regions of ($E,$ $l^{2}$) where the phases are clearly defined. In
addition, the time averaged virial ratio of the system prepared in the
transition and high energy regions were found to converge on a time scale
related to the convergence of the one particle probability density function
in $\mu $ space described by Eq.~(\ref{relax}). This is in contrast with the
behavior below the transition region where, typically, the system rapidly
''virialized'' in about 100 dynamical times. At energies above the
transition region, $\overline{VR}$ approached the equilibrium value through
a sequence of ''under-damped'' oscillations, while below this region the
evolution was characteristic of either critical or over-damping which shows
that mixing is much more effective at low energies.

In common with the earlier study of irrotational shells \cite{millyounglett}%
, we selected the virial ratio as a useful order parameter. In Fig.~\ref
{figphasesim}, we plot the time averaged virial ratios of the relaxed
systems for different values of the energy for 56 simulation runs with $%
l^{2}=5\times 10^{-5}$. As we can see, the simulation results appear to
converge to the mean field predictions with increasing $N$, and the
transition region is broadened due to finite size effects. It is interesting
that, above the transition, the ''experimental'' virial ratio is always less
than the mean field prediction for any N while, below the transition point,
the behavior is reversed. We would expect this behavior if the system were
spending varying amounts of time in each phase, and we will discuss this
possibility further in the conclusions. Moreover, above the transition
region, the approach to mean field behavior with increasing $N$ (not time)
occurs more slowly than below the transition energy. This is consistent with
the observation that, at higher energies, the inner part of the relaxed
density profiles are greater than the mean field prediction in the central
region. This effect is most evident for $N=16, 32$ (Fig.~\ref{figdensityhigh}%
).

Phase transitions are only sharp in the limit $N\rightarrow \infty $ . For
finite $N$ the transition point is shifted and the sharpness is ''rounded''.
In ''normal'' systems with short range interactions, it has been proved that
the both amount of shifting and rounding scale with distinct powers of $N$ 
\cite{pathria}. We carefully verified that the shifting of the transition
point and the rounding of the transition region is in very good agreement
with finite size scaling theory. We found the transition energy satisfies $%
E_{tr}\left( N\right) -E_{tr}\left( \infty \right) \propto N^{-\lambda }$
while the width of the transition region scales as $\Delta E\propto
N^{-\theta }$ \cite{binder,pathria} with shifting and rounding exponents
given by $\lambda =1.42$ and $\theta =1.02$ , respectively. This result
shows that finite size effects in gravitational systems can also be
explained by scaling theory. Finite size scaling was also confirmed for the
system of irrotational shells \cite{millyounglett,youngmill2}.

\section{Fluctuations and Correlations}

While it is useful to compare the time averages of physical quantities
obtained from simulations with the predictions of equilibrium mean field
theory, dynamical simulations also provide an opportunity to investigate
other system properties which are not addressed by MFT, such as the average
size of fluctuations and the decay of correlations in both position and
time. In our simulations we continuously monitored the kinetic energy. For
the smallest value of $N$, the spontaneous fluctuations were large, on the
order of the mean value, but for the larger populations they settled down.
As a further check on the convergence to MFT, we studied the population
dependence of the variance of statistical fluctuations in both the kinetic
energy and the population of each bin. In the Vlasov limit the system is
completely described by the single particle distribution $f(r,v)$ . If this
description is valid, i.e. if the system is approaching the Vlasov limit
with increasing $N$, then the variances should be asymptotically decreasing
as $1/N$ for sufficiently large $N$. It is interesting that, even for the
limited range of total system population considered here, the decay of the
bin populations obeyed this law almost exactly. The situation for the
fluctuations in kinetic energy was not so simple. While their variance also
decayed with increasing $N$, the observed rate of decrease was not uniformly
proportional to $1/N$, but rather depended strongly on which part of the ($%
E, $ $l^{2}$) phase plane the system was situated. We will return to this
point later.

Useful information concerning the system dynamics and the approach to the
Vlasov limit can be gleaned from an examination of both correlations in time
Eq.~(\ref{KEcorr}) and position Eq.~(\ref{popcorr}). Since there is no
information loss in true Vlasov dynamics, to the extent that this
description is accurate, the duration of correlations in time and,
correspondingly, the range of correlations in position, may not decay to
zero. This type of behavior has been observed previously in Vlasov
simulations of the one dimensional system of planar mass sheets, where
complex structures in the $\mu $ space distribution appeared to persist
indefinitely \cite{mineufeix}. Following common practice, notice that in our
definition of the correlation functions we have normalized them to unity at $%
t=0$ for the kinetic energy, and on the diagonal for bin populations. Thus
differences of their values from unity reflect the duration and range of
correlation. In Table~\ref{table_1}, we list the values of energy and $l^{2}$
which were used in our simulations in each region of the phase plane. Since
the transition point is shifted as a result of finite size effects, for the
evaluation of the population correlation matrix in Eq.~(\ref{popcorr}) at
this point, we used the equal probability bin radii derived from the high
energy mean-field density profile. In practice, using the low energy bin
radii did not have any impact on the population correlation matrix at the
transition point.

The duration of correlation in the total kinetic energy of the system
provides a useful indicator for the lifetime of fluctuations of macroscopic
quantities. In Fig.~\ref{figKEcorr}, we plot the kinetic energy time
correlation function in each of the five different phase regions for $N=64$.
In general we observed that correlations in the supercritical and high
energy phases decay rapidly, on the order of a few dynamic times. In
contrast, the duration of correlation in the low energy and supercritical
phases was at least an order of magnitude longer. In the transition region
there appears to be a shift at approximately 15 dynamical times to a much
slower decay which is hard to quantify from the graph, but may still be
present after 100 dynamical time units. The same qualitative behavior was
observed for each value of the population studied. However the duration of
correlation in each phase is increasing with increasing $N$. This is
consistent with other dynamical studies of systems with long range
interaction, which show that Lyapunov exponents decrease, and hence memory
effects endure, with increasing population \cite{toshio} once a critical
value of $N$ has been exceeded \cite{MillerReidl}. We also mention that, at
the low energy of $E=1$ with populations $N=32,64,128$, correlations started
becoming smaller and similar to those at high energy, while we observed
significant correlations of long duration in the supercritical phase at much
larger values of $l^{2}$. This will be discussed in the following sections.

Correlations in position are reflected by nonvanishing off-diagonal elements
of the correlation matrix. In Figures \ref{figpopcorrlow}, \ref
{figpopcorrphase}, and \ref{figpopcorrcrit}, we present the population
correlation matrix $C_{ij}$ with $N_{bin}=20$ and $N=64$ for the low energy
phase ($E=1$), transition region ($E=2$), and at the critical point. In the
high energy region above the transition, and in the supercritical region for
small values of $l^{2}$, no significant correlations were present in the
system. In the low energy phase (Fig.~\ref{figpopcorrlow}), strong
correlation is only present near the system center where the density is
high. This effect may be due to the presence of the central core, or long
lasting collective oscillations, which we discuss below. Note that the
anti-correlated domain in Fig.~\ref{figpopcorrlow} coincides with the region
where the central density decreases to the dilute halo background in the
mean-field density profile, at about bins $3-5$. In the transition region
however (\ref{figpopcorrphase}), long range, correlations are clearly
present. The likely explanation is found by inspection of Fig.~\ref
{figKEcorr}, where we found an extremely long time tail in the kinetic
energy auto-correlation function. It appears that, close to the transition,
slowly decaying diffusive modes are propagating throughout the entire
system. Effectively, the system may be spending some time in each phase.
This simple idea would explain many of the observations for the transition
region including the system-wide correlation, and the long time tail in the
autocorrelation function, and will be discussed in the conclusions. For $%
N=128$, this effect is more evident because the transition region is less
rounded. Near the critical point (Fig.~\ref{figpopcorrcrit}), we observe
similar behavior as in the low energy phase, except that the correlated
region is broadened. Here we also confirmed that the anti-correlated region
in Fig.~\ref{figpopcorrcrit} coincides with the region where the mean-field
density fades into the halo background, which occurs at about bins $3-8$.
Note that oscillations in the kinetic energy autocorrelation function also
appear here.

\begin{table}[tbp]
\caption{The energy and $l^{2}$ values used in the simulations for the
corresponding phases.}
\label{table_1}%
\begin{ruledtabular}
\begin{tabular}{clc}
Type of phase & Energy & $l^{2}$\\
\hline
High energy phase & 4 & $5\times 10^{-5}$\\
Low energy phase & 1.5 & $5\times 10^{-5}$\\
Low energy phase & 1 & $5\times 10^{-5}$\\
Mean-field transition point & 1.9 & $5\times 10^{-5}$\\
Transition region & 2 & $5\times 10^{-5}$\\
Fluid phase & 4 & $5\times 10^{-3}$\\
Critical point & 1.052 & $1.1\times 10^{-4}$\\
\end{tabular}
\end{ruledtabular}
\end{table}

\section{Relaxation}

In this work we consider three types of relaxation, violent relaxation
following the initial phase of the simulation, equilibration, i.e. the
approach to equilibrium, on the longest time scales, and the decay of
kinetic energy fluctuations once equilibrium has been obtained. We
investigate early (violent) relaxation by studying the decay of oscillations
in the averaged virial ratio $\overline{VR}$, equilibration by the reduction
of the $\sigma _{r}^{2}$ statistic Eq.~(\ref{relax}) with time, and the
decay of fluctuations through the kinetic energy auto-correlation function
Eq.~(\ref{KEcorr}). We see below that each type of relaxation has
characteristics which reflect the location in the $(E,l^{2})$ phase plane in
which the system is prepared.

In the initial evolutionary stage, both the kinetic and potential energy are
changing rapidly in time. Lynden-Bell termed this stage ''violent
relaxation'' and pointed out that phase mixing is the usual mechanism for
the establishment of a quasi-stationary distribution in $\mu $ space which
causes the gravitational system to virialize. Early relaxation studies of
the planar sheet model showed that virialization typically takes place in
about 50 dynamical times, or less \cite{hohlbroadus}. Here we studied the
development of the time averaged virial ratio for each of the initial states
discussed earlier as time progressed. Again we found a varied behavior which
depends strongly on the location of the initial $(E,l^{2})$ phase point.
Virialization occurred most rapidly in the low energy phase (Fig.~\ref{figvirrelaxlow}) 
in about $100t_{dyn}$. On the other hand, in the high energy phase region, the process of virialization takes place in a much longer time scale (Fig.~\ref{figvirrelaxhigh}). 
In common with earlier studies, for both the high and low energy
phases, the time to achieve virialization decreased with increasing $N$. In
contrast, in the transition region, virialization was similar to the
relaxation process in $\mu $ space (Fig.~\ref{figvirrelaxphase}) and takes
place on a much longer time scale. Moreover, with increasing $N$,
virialization took place on longer time scales due to the decrease in
broadening of the transition region, and the increasing influence of the
metastable state (Fig.~\ref{figvirrelaxphase}).

The $\sigma _{r}^{2}$ statistic Eq.~\ref{relax} compares the time averaged
distribution of shells in bins at times $t$ and $2t$. As noted earlier, it
was computed for all initial conditions, and its value was used to terminate
each run. Ideally, if the system has perfectly equilibrated, it should
vanish. In Figures \ref{figrelaxhigh}, \ref{figrelaxphase}, and \ref{figrelaxlow}, 
we present log-log plots of $\sigma_{r}^{2}$ versus time in the three ditinct phase regions for simulations with 
$N=16, 32, 64, 128$. In order to compensate for the statistical error induced by
differences in $N$, we used sampling rates with respective snapshot times 
$0.125t_{dyn}, 0.25t_{dyn}, 0.5t_{dyn}, t_{dyn}$. In the transition region, we
used the high energy equal mass mean-field bins to calculate the relative
populations. In all cases we found that the relaxation process exhibits
power law behavior on long time scales, which means $\sigma _{r}^{2}\propto
t^{-\alpha }$. Interestingly, for both the high (Fig.~\ref{figrelaxhigh}) and low energy (Fig.~\ref{figrelaxlow}) phases, the exponent $\alpha $ was not $N$ dependent, and the relaxation was similar in both phases with $\alpha =1.0\pm 0.1$. However, in the transition region
simulations were relaxing more slowly with larger fluctuations (Fig.~\ref{figrelaxphase}), suggesting
that the system may be flipping back and forth between the stable and
metastable phases. Here, with increasing $N$, $\alpha $ became smaller while
its variance becomes larger. For $N=16$, $\alpha $ is still $1\pm 0.2$ but
for $N=32, 64, 128$ the uncertainty becomes increasingly larger.

The possibility for collective modes in systems with long range forces has
been studied in both plasmas and gravitational systems \cite{binney}. Even
for the small populations considered here, careful examination of Fig.~\ref
{figKEcorr} indicates that, as the energy is lowered, the correlation
function exhibits damped oscillations. Since we are examining the total
kinetic energy of the system, this suggests the possible presence of
collective oscillations in equilibrium. To examine this in further detail,
we used $C_{KE}\left( \tau \right) $ to study the decay of kinetic energy
fluctuations in the fluid phase with fixed $E=0.0$ for different values of
both $l^{2}$ and $N$. In Fig.~\ref{figKEcorrfluidl2}, we present the kinetic
energy auto-correlation function for fixed $E=0$ and three values of $l^{2}$
($0.005,0.1,0.5$) above the critical point value. In a system with
sufficiently large $l^{2}$, we see that oscillations start to develop with a
long-lived positive tail. In Fig.~\ref{figKEcorrfluidN}, we plot the
correlation functions for different values of $N$ at $E=0$ and $l^{2}=0.1$.
Note that as we approach the mean-field limit, the non-vanishing tail has a
larger value, and the oscillating part has a longer duration. This suggests
that, with increasing $N$, single particle behavior is suppressed and the
system starts to behave like a Vlasov fluid. In the mean-field limit, the
effective potential $\Phi +l^{2}/(2r^{2})$ always has a minimum. For large
values of $l^{2}$, the width of this region becomes larger, possibly
allowing low frequency collective modes to develop in the system. Similar
behavior was observed by Rouet et. al. for the system of planar mass sheets 
\cite{rouet}. There the initial system was prepared in a stationary
''waterbag'' state. They found that kinetic energy fluctuations had the
frequency of the collective modes, rather than that of the individual closed
Vlasov orbits in the ($x,$ $v)$ phase plane. The same phenomena may be
occurring in the $l^{2}$ model as well. Supporting evidence was provided by
the correlation matrix which showed strong correlations in position in the
central part of the system for these thermodynamics states.

\section{Summary and Conclusions}

The role of thermodynamics in controlling the evolution of gravitational
systems is only partially understood, and there are many open questions. It
is our impression that recently this subject is attracting increased
attention. Observations of the radial density dependence of globular
clusters show that they fall into two groups, with either a smoothly
decreasing density profile with increasing radius, $r$, or a sharp central
peak and a more diffuse halo \cite{Heggie}. The evidence suggests that a
thermodynamic interpretation may be possible, i.e. perhaps the clusters can
exist in different phases at observable times \cite{Heggie}.

Here we investigated the dynamics of the gravitational phase transition in
the $l^{2}$ model of a self-gravitating system in the MCE in which the mass
elements are thin, concentric, shells. In this idealized model, instead of
regularizing the singularity of the Newtonian potential, we simply fixed the
square of the angular momentum of each shell. In section \ref{Nbody}, we
showed that, in the mean field limit, the equilibrium states of this system
correspond to those of the more realistic system of point masses when the
same constraint is imposed. In an earlier work, we showed that a
correspondence also exists under less restrictive circumstances \cite
{klinkmillL2}. The constraint of constant $l^{2}$ for each shell, or
particle, establishes a centrifugal barrier which prevents the occurrence of
the gravothermal catastrophe, and induces a first order phase transition in
both CE and MCE \cite{klinkmillprok}.

One important conclusion which can be reached from this study is that
angular momentum exchange plays an important role in a self-gravitating
system which alters the thermodynamic behavior. Another is that it is not
necessary to soften the singularity in the Newtonian potential to obtain a
transition. The effect of the singularity can be blocked by other
mechanisms. This is especially relevant in stellar clusters, where the
distance between stars is too great for softening to be important. If
stellar systems can exist, or even approximate, different thermodynamic
phases, the influence of the singularity needs to blocked, at least
temporarily. In our earlier, mean field, study of a spherical system which
permitted exchange, no transition was found and the gravothermal catastrophe
could not be prevented in spite of the fact that both $L_{2}$ and $E$ were
fixed. At this time it is not clear how this might occur. One possibility is
that angular momentum exchange between stars may occur so slowly that, in
the present epoch, thermodynamics could be influenced by an effective
centrifugal barrier. Since globular clusters are approximately spherical,
this is worth investigating. Another possibility is that stellar
interactions with hard binaries in the cluster core may also establish a
centrifugal barrier. These conjectures require further investigation.

In this work we used N-body dynamical simulation to verify the earlier
mean-field predictions \cite{klinkmillprok}. Our N-body simulations
confirmed the mean-field phase transition, which was shifted and broadened
due to finite size effects. We also verified that finite size scaling is in
very good agreement with the observed shifting and rounding in the
transition region. An interesting feature of the convergence to MFT was the
observation that agreement in the density with increasing $N$ was much
slower near the system center for small values of $l^{2}$. This appears to
be a discreteness effect. For a small value of $l^{2}$ the density is
changing rapidly due to the competition between the centrifugal barrier and
the largely unscreened gravitational potential, which combine to form a very
narrow minimum in the effective potential. Because of discreteness effects,
the mean local density cannot change this rapidly in the dynamical
simulation, and this was readily observable in the average population of the
central bin.

In addition to confirming agreement with mean field theory for the time
average of physical quantities, such as the density and the kinetic and
potential energies, we also studied fluctuations in density and kinetic
energy, correlations in position, and the dynamical behavior of the system
in each phase, the transition region, and at the critical point, i.e.
properties which are not accessible from MFT. These included virialization,
relaxation to equilibrium, and the decay of fluctuations in kinetic energy
through its autocorrelation function. In general, as $N$ becomes large, it
is expected that the Vlasov regime will be approached. In this regime $%
f(r,v) $ completely characterizes the system at each time. It was shown long
ago that in a system with a smooth, bounded, potential, fluctuations and
correlations decay in the Vlasov limit \cite{braunhepp}. In the large $N$
scaling regime it is easy to show that the variance of fluctuations in both
bin populations and total kinetic energy should decay as $N^{-1}$. We found
this behavior over the complete range of population for the former, while a
wide variation of power law exponents characterized the reduction of the
kinetic energy variance depending on the thermodynamic state. This suggests
that our simulations were not fully in the scaling regime for the smaller
populations considered. This is not surprising, and was supported by the
size of the spontaneous kinetic energy fluctuations as well as the $N$
dependence of the remaining dynamical quantities.

In the transition region, strong corroborating evidence was obtained to
support the ansatz that the system is fluctuating in time between the stable
and metastable phase. First of all, relaxation to equilibrium as measured by
the $\sigma _{r}^{2}$ statistic takes longer and, at a given time, the
fluctuations are much larger than those occurring away from the transition.
Second, virialization occurs on the same time scale as $\sigma _{r}^{2}$,
whereas at low energy, the virialization takes place in about $100t_{dyn}$.
Fast virialization is also observed in other model systems which lack a
transition, such as the planar sheet system\cite{hohlbroadus,toshio2}, and
point mass systems \cite{binney, Heggie} where it occurs in 50-100 dynamical
times, long before relaxation to equilibrium has occurred. Third, near the
transition an extremely long time tail appears in the kinetic energy
autocorrelation function, indicating that the system continues to feel the
presence of the metastable phase. The lack of oscillation in $C_{KE}(t)$
suggests that a slow, diffusive, mode is dominating the linear relaxation.

The observation that relaxation to equilibrium exhibits power law behavior
was unanticipated and is crucial for understanding the system dynamics. In
the Astrophysics literature it is usual to define a relaxation time for
gravitational evolution \cite{Heggie,binney} as the time it takes for a
typical star to be sufficiently deflected that the change in its velocity is
on the order of its mean. The standard result is $t_{relax}\approx (N/10\ln
N)t_{dyn}$. However, here we see that relaxation does not occur with a
characteristic time, but rather is scale free, with $\sigma _{r}^{2}\sim
t^{-1}$. An open question is whether relaxation is scale free in higher
dimension as well, e.g. for a point mass system.

There are now two gravitational systems in which the existence of a phase
transition has been demonstrated both by mean field theory and, to within
finite scaling, dynamical simulation. The system of irrotational shells
studied earlier \cite{millyounglett} and the system considered here share
many similarities, but differ in one feature. Each system is characterized
by an additional parameter besides the energy: In the non-rotating model,
the inner barrier radius $a$ plays the role of $l^{2}$ so its maximum
density occurs at $r=a$. Each of these systems exhibits a transition in MCE
to a more centrally condensed phase as the energy is lowered. Each system
has a critical value of the new parameter above which the transition is not
allowed. However, in the present system we find evidence of persistent
fluctuations that develop long lived oscillations as $N$ is increased in
MCE, and local correlation in position near the system center. This suggests
that collective oscillations may be occurring in the system. Persistent
oscillations were also observed in simulations of the planar sheet system
for large $N$ \cite{rouet}.

We are currently modelling the system via dynamical simulation for the
canonical ensemble. This is accomplished by introducing thermalizing
collisions at the outer boundary. There are strong contrasts with the MCE
which we will report in a separate work. We also plan to use dynamical
simulation to investigate the locally stable regime which arises when
angular momentum exchange is allowed \cite{klinkmillL2}. In addition, we
would like to suggest that Vlasov dynamics may give insight into the
collective behavior found in the fluid phase of the $l^{2}$ model.

\bibliographystyle{apsrev}

\end{document}